\documentclass[showpacs,preprintnumbers,11pt]{revtex4}
\usepackage{bm}

\begin{document}

\title{Quantum Statistical Relation for black holes in nonlinear
electrodynamics coupled to Einstein-Gauss-Bonnet AdS gravity}
\author{Olivera Miskovic}
\email{olivera.miskovic@ucv.cl}
\affiliation{Instituto de F\'{\i}sica, Pontificia Universidad Cat\'{o}lica
de Valpara\'{\i}so,\\
Casilla 4059, Valpara\'{\i}so, Chile}
\author{Rodrigo Olea}
\email{rodrigo.olea@ucv.cl}
\affiliation{Instituto de F\'{\i}sica, Pontificia Universidad Cat\'{o}lica de
Valpara\'{\i}so,\\
Casilla 4059, Valpara\'{\i}so, Chile}
\date{\today}

\begin{abstract}
We consider curvature-squared corrections to Einstein-Hilbert gravity action
in the form of Gauss-Bonnet term in $D>4$ dimensions. In this theory, we
study the thermodynamics of charged static black holes with anti-de Sitter
(AdS) asymptotics, and whose electric field is described by nonlinear
electrodynamics (NED). These objects have received considerable attention in
recent literature on gravity/gauge dualities.

It is well-known that, within the framework of anti de-Sitter/Conformal
Field Theory (AdS/CFT) correspondence, there exists a nonvanishing Casimir
contribution to the internal energy of the system, manifested as the vacuum
energy for global AdS spacetime in odd dimensions. Because of this reason,
we derive a Quantum Statistical Relation directly from the Euclidean action
and not from the integration of the First Law of thermodynamics. To this
end, we employ a background-independent regularization scheme which consists
in the addition to the bulk action of counterterms that depend on both
extrinsic and intrinsic curvatures of the boundary (Kounterterm series).
This procedure results in a consistent inclusion of the vacuum energy and
chemical potential in the thermodynamic description for
Einstein-Gauss-Bonnet AdS gravity regardless the explicit form of the NED
Lagrangian.
\end{abstract}

\pacs{04.50.-h, 04.50.Gh, 04.70.Dy, 11.30.-j}
\maketitle

%%%%%%%%%%%%%%%%%%%%%%%%%%%%%%%%%%%%%%%%%%%%%%%%%%%%%%%%%%%%%%%%%%%%%%%%%%%%%%%%%%%%%
%%%%%%%%%%%%%%%%%%%%%%%%%%%%%%%%%%%%%%%%%%%%%%%%%%%%%%%%%%%%%%%%%%%%%%%%%%%%%%%%%%%%%

\section{Introduction}

AdS/CFT correspondence \cite{AdS/CFT} provides a computational tool to
obtain physical properties of a field theory at strong coupling by studying
its gravity dual.

In hydrodynamic models, for example, gauge/gravity duality is used to
calculate the ratio of shear viscosity $\eta $ to entropy density $s$ in a
strongly interacting quantum theory. In particular, it predicts a universal
bound $1/4\pi $ of this ratio for a large class of theories, dual to
Einstein AdS gravity \cite{KSS}.

However, it has been shown that the addition of higher-derivative terms to
the gravitational action in AdS space modifies the dynamics of the boundary
theory and it violates the KSS bound \cite{viscosity}. On the other hand,
higher-curvature terms in Abelian fields described by nonlinear
electrodynamics (NED) do not change the value $\eta /s$ \cite{Cai-Sun},
unless gravity action is augmented by a Gauss-Bonnet (GB) term, what also
causes instabilities in the large-momentum regime \cite%
{Ge-Matsuo-Shu-Sin-Tsukioka}.

In a similar fashion, when studying gravity duals to high critical
temperature superconductivity models, the inclusion of GB and NED terms
alters the ratio between $T_{c}$ and the energy gap \cite%
{Gregory-Kanno-Soda,Pan-Wang-Papantonopoulos-Oliveira-Pavan,Jing-Chen}.

In the examples of AdS/CFT duality mentioned above, the mapping between the
bulk and boundary theories makes an extensive use of thermal properties of
black holes in the gravity side of the correspondence. This provides the
motivation to investigate the thermodynamics of black holes in
Einstein-Gauss-Bonnet (EGB) gravity with negative cosmological constant
coupled to an arbitrary NED theory. In particular, we prove here that any
black hole which is a solution to the theory will satisfy the Quantum
Statistical Relation (QSR) \cite{footnote1}%
\begin{equation}
T\,I^{E}=U-T\,S+\Phi Q\,,  \label{QSR1}
\end{equation}%
where $T$ is the Hawking temperature, $U$ is the black hole energy and $S$
its entropy that includes a correction due to GB term. The addition of a NED
Lagrangian to pure gravity action brings in electric charge $Q$ to the
solution, whose conjugate variable $\Phi $ is the difference in electric
potential between the horizon and infinity. In other words, the Euclidean
action is the Legendre transformation of the entropy with respect to
chemical (electric) potential. As it is standard within the framework of
AdS/CFT, the Euclidean action has to be regularized by means of a
background-independent procedure.

In gravity with AdS asymptotics, there is a subtle difference between the
relation (\ref{QSR1}) and the First Law of thermodynamics%
\begin{equation}
dU=TdS-\Phi dQ\mathcal{\,},  \label{First Law}
\end{equation}%
because $U$ contains in general a Casimir contribution in $D=2n+1$
dimensions \cite{footnote2}. Clearly, the latter formula remains the same if
the total energy is shifted by an additive constant $E_{vac}$. This does not
mean that the presence of a vacuum energy is irrelevant. On the contrary, it
is expected that $E_{vac}$ will appear in a consistent thermodynamic
description of asymptotically AdS (AAdS) black holes, when we work out their
thermal properties from the partition function in semiclassical
approximation
\begin{equation}
Z=e^{-I_{clas}^{E}}.  \label{Z}
\end{equation}%
Here, $I_{clas}^{E}$ stands for the classical Euclidean action, which is
divergent in AdS gravity. If one intends to identify any thermodynamic
potential with the Euclidean action, one must resort to a regularization
mechanism that eliminates the divergences in the infrared sector of the
theory, but not the vacuum energy. It is only in this way that we can
extract $I^{E}$ as the finite part of $I_{clas}^{E}$, and relate it to the
asymptotic charges that appear in Eq.(\ref{QSR1}).

In the context of AdS/CFT correspondence, the regularization of the
Euclidean action requires the addition of counterterms which are intrinsic
functionals of the boundary metric and curvature \cite%
{Balasubramanian-Kraus,Emparan-Johnson-Myers}, and constructed using
holographic techniques for AAdS spacetimes \cite%
{Henningson-Skenderis,deHaro-Skenderis-Solodukhin}. When a GB term is
included in the gravity action --despite the fact the field equations are
still second-order in the metric-- one faces almost unsurmountable obstacles
to apply the holographic renormalization procedure. Thus, it is necessary to
postulate the explicit form of the local counterterm series and, then, the
coefficients in the series are adjusted by the convergence of charges and
Euclidean action for particular solutions \cite{Brihaye-Radu,Liu-Sabra}.
This seems to work well for low enough dimensions, but it is far from giving
a general prescription for all cases.

In this paper, we apply an \textit{extrinsic}, background-independent
regularization method for EGB AdS gravity to render the Euclidean action
finite in all dimensions \cite{Kofinas-Olea-GB}. This scheme involves the
addition to the action of counterterms depending on both the intrinsic and
extrinsic curvatures (which, for that reason, have been suggestively called
\textit{Kounterterms}). This procedure results in the obtention of a Quantum
Statistical Relation for black hole solutions to Einstein-Gauss-Bonnet AdS
gravity coupled to an arbitrary NED. As we shall show below, the picture in
odd dimensions is consistent only if the total energy is shifted as $%
U=M+E_{vac}$ with respect to the \textit{Hamiltonian} mass $M$, in a similar
fashion as for Einstein-Born-Infeld case studied in Ref.\cite%
{Miskovic-OleaBI}.

\section{Field equations}

Nonlinear actions for electromagnetic field have been a subject of research
for many years, since it was noted by Heisenberg and Euler that quantum
corrections to electrodynamics lead to nonlinear equations for the field
strength \cite{Heisenberg-Euler}. The physics of objects described by
nonlinear effective Lagrangians, as the one of Born and Infeld \cite%
{Born-Infeld}, possesses remarkable properties, such as a regular electric
field at the origin and a finite self-energy. Furthermore, Born-Infeld
electrodynamics can be obtained in exact one-loop computations from open
Bose strings \cite{Fradkin-Tseytlin}.

Gravity coupled to NED leads to a large class of charged black hole
solutions. In Einstein-Hilbert gravity coupled to Born-Infeld
electrodynamics, static, spherically symmetric black holes were derived in,
e.g., Refs. \cite{Hoffman-Gibbons-Rasheed,Oliveira}. Among other gravitating
NED models that exhibit electrically charged solutions, we can mention the
logarithmic theory investigated in \cite{Soleng}, and the conformally
invariant ED found in \cite{Maeda-Hassaine-Martinez}.

We will study gravitating NED in any spacetime dimension $D>4$ which is
described by the action
\begin{equation}
I_{0}=\int\limits_{\mathcal{M}}d^{D}x\,\sqrt{-g}\,\mathcal{L}%
_{0}=I_{grav}+I_{NED}\,.  \label{bulk action}
\end{equation}%
The first part of the bulk action corresponds to the one of Einstein-Hilbert
(EH) with negative cosmological constant $\Lambda =-\left( D-1\right) \left(
D-2\right) /2\ell ^{2}$ and a higher-curvature correction given by the
Gauss-Bonnet term
\begin{equation}
I_{grav}=\frac{1}{16\pi G}\int\limits_{\mathcal{M}}d^{D}x\,\sqrt{-g}\,\left[
R-2\Lambda +\alpha \,\left( R^{2}-4R_{\mu \nu }R^{\mu \nu }+R_{\mu \nu
\lambda \sigma }R^{\mu \nu \lambda \sigma }\right) \right] \,.
\label{EGB AdS}
\end{equation}%
Here, $\ell $ is the AdS radius, $G$ is the gravitational constant and $%
\alpha $ is the GB coupling, which has to be positive if one regards this
action as obtained in the low-energy limit of String Theory.

The Abelian gauge field $A_{\mu }(x)$ defines a field strength $F_{\mu \nu
}=\partial _{\mu }A_{\nu }-\partial _{\nu }A_{\mu }$, which couples
minimally to gravity through the quadratic invariant $F^{2}=g^{\mu \lambda
}g^{\nu \rho }F_{\mu \nu }F_{\lambda \rho }$. We consider an arbitrary
Lagrangian density $\mathcal{L}(F^{2})$, such that the NED action has the
form
\begin{equation}
I_{NED}=\int\limits_{\mathcal{M}}d^{D}x\,\sqrt{-g}\,\mathcal{L}(F^{2})\,.
\label{NED}
\end{equation}

The equations of motion follow from the variation of the bulk action respect
to the dynamic fields $g_{\mu \nu }$ and $A_{\mu }$, that is, $\delta
I_{0}/\delta g_{\mu \nu }=0$ and $\delta I_{0}/\delta A_{\mu }=0$,%
\begin{eqnarray}
\mathcal{E}_{\nu }^{\mu } &\equiv &G_{\nu }^{\mu }+H_{\nu }^{\mu }-8\pi
G\,T_{\nu }^{\mu }=0\,,  \label{E^mu_nu} \\
\mathcal{E}^{\mu } &\equiv &\nabla _{\nu }\left( F^{\mu \nu }\frac{d\mathcal{%
L}}{dF^{2}}\right) =0\,.  \label{E_mu}
\end{eqnarray}%
In Eq.(\ref{E^mu_nu}), $G_{\nu }^{\mu }$ is the Einstein tensor which
includes the contribution from the cosmological constant%
\begin{equation}
G_{\nu }^{\mu }=R_{\nu }^{\mu }-\frac{1}{2}\,\delta _{\nu }^{\mu }R+\Lambda
\,\delta _{\nu }^{\mu }\,,  \label{Einstein tensor}
\end{equation}%
whereas $H_{\nu }^{\mu }$ denotes the Lanczos tensor
\begin{eqnarray}
H_{\nu }^{\mu } &=&-\frac{\alpha }{2}\,\delta _{\nu }^{\mu }\left(
R^{2}-4R^{\alpha \beta }R_{\alpha \beta }+R^{\alpha \beta \lambda \sigma
}R_{\alpha \beta \lambda \sigma }\right)  \nonumber \\
&&+2\alpha \left( RR_{\nu }^{\mu }-2R^{\mu \lambda }R_{\lambda \nu
}-2R_{\lambda \nu \sigma }^{\mu }R^{\lambda \sigma }+R^{\mu \alpha \lambda
\sigma }R_{\nu \alpha \lambda \sigma }\right) \,,  \label{Lanczos tensor}
\end{eqnarray}%
which comes from the GB part of the gravity action. The right hand side of
Eq.(\ref{E^mu_nu}) corresponds to the matter stress tensor $T^{\mu \nu }=%
\frac{2}{\sqrt{-g}}\frac{\delta I_{NED}}{\delta g_{\mu \nu }}$ which has the
form%
\begin{equation}
T_{\nu }^{\mu }=\delta _{\nu }^{\mu }\,\mathcal{L}-4\,\frac{d\mathcal{L}}{%
dF^{2}}\,F^{\mu \lambda }F_{\nu \lambda }\,.
\end{equation}%
For a given solution ansatz, the generalized Maxwell equation (\ref{E_mu})
determines the dynamics of the electromagnetic field.

The presence of the higher-curvature term $H_{\nu }^{\mu }$ in the equation
of motion (\ref{E^mu_nu}) changes the vacua structure of AdS gravity.
Indeed, a nonvanishing GB coupling $\alpha $ modifies the AdS radius $\ell $
of maximally symmetric spaces to an effective value $\ell _{e\!f\!f}$ given
by%
\begin{equation}
\ell _{e\!f\!f}^{(\pm )2}=\frac{2\alpha \left( D-3\right) \left( D-4\right)
}{1\pm \sqrt{1-\frac{4\alpha }{\ell ^{2}}\,\left( D-3\right) \left(
D-4\right) }}\,,\qquad \alpha \leq \frac{\ell ^{2}}{4\left( D-3\right)
\left( D-4\right) }\,.  \label{l_eff pm}
\end{equation}%
A multiplicity of solutions may exist in either branch of the theory, some
of them yet to be discovered. For the present paper we will consider
spacetimes which are AAdS. In order to impose this condition in a way
independent of any coordinate frame, we will assume the AAdS behavior in the
curvature rather than in the metric, that is,
\begin{equation}
R_{\mu \nu }^{\alpha \beta }\rightarrow -\frac{1}{\ell _{e\!f\!f}^{2}}%
\,\delta _{\left[ \mu \nu \right] }^{\left[ \alpha \beta \right] }\,,
\label{LAdSleff}
\end{equation}%
for the corresponding branch.

For static black holes with constant-curvature transversal section $\Gamma
_{D-2}$, described by the metric $\gamma _{nm}$, the line element in the
local coordinates $x^{\mu }=(t,r,\varphi ^{m})$ reads%
\begin{equation}
ds^{2}=g_{\mu \nu }(x)\,dx^{\mu }dx^{\nu }=-f^{2}(r)\,dt^{2}+\frac{dr^{2}}{%
f^{2}(r)}+r^{2}\gamma _{mn}(\varphi )\,d\varphi ^{m}d\varphi ^{n}\,.
\label{BH}
\end{equation}%
The curvature of $\Gamma _{D-2}$ is labeled by the topological parameter $%
k=0,$ $+1$ or $-1$ for flat, spherical or hyperbolic case, respectively. The
spacetime boundary $\partial \mathcal{M}$ has the topology of a cylinder and
is placed at $r\rightarrow \infty $. A necessary condition for a solution to
be a black hole is the existence of a horizon $r_{+}$ such that $%
f^{2}(r_{+})=0$. In case the latter relation possesses more than one root, $%
r_{+}$ denotes the outermost one.

A charged static solution is given in terms of a gauge field which depends
only on the radial coordinate%
\begin{equation}
A_{\mu }=\phi \left( r\right) \,\delta _{\mu }^{t}\,,  \label{A}
\end{equation}%
what generates the electric field
\begin{equation}
E(r)=-\phi ^{\prime }(r)\,,  \label{def_E}
\end{equation}%
such that the field strength is%
\begin{equation}
F_{\mu \nu }=E(r)\,\left( \delta _{\mu }^{t}\delta _{\nu }^{r}-\delta _{\nu
}^{t}\delta _{\mu }^{r}\right) \,.  \label{F_ansatz}
\end{equation}%
Here, the prime stands for radial derivative.

In the static ansatz (\ref{BH}--\ref{F_ansatz}), the gravitational equation (%
\ref{E^mu_nu}) gives rise to a single independent differential equation for $%
f^{2}$,%
\begin{eqnarray}
\mathcal{E}_{t}^{t} &=&\mathcal{E}_{r}^{r}=\frac{D-2}{2r^{2}}\,\left[
r\left( f^{2}\right) ^{\prime }+\left( D-3\right) \left( f^{2}-k\right)
-\left( D-1\right) \frac{r^{2}}{\ell ^{2}}\right]  \nonumber \\
&&+\alpha \,\left( D-2\right) \left( D-3\right) \left( D-4\right) \,\frac{%
k-f^{2}}{r^{3}}\,\left[ \left( f^{2}\right) ^{\prime }-\left( D-5\right) \,%
\frac{k-f^{2}}{2r}\right] -8\pi G\,T_{r}^{r}\,,  \label{grav diff Eq.}
\end{eqnarray}%
where\textbf{\ }%
\begin{equation}
T_{t}^{t}=T_{r}^{r}=\left. \left( \mathcal{L}+4E^{2}\frac{d\mathcal{L}}{%
dF^{2}}\right) \right\vert _{F^{2}=-2E^{2}}\,.  \label{T components}
\end{equation}

On the other hand, the form of the electric field (\ref{F_ansatz}) leads to
the differential version of a generalized Gauss' law
\begin{equation}
\mathcal{E}^{t}=-\frac{d}{dr}\,\left( r^{D-2}E\,\left. \frac{d\mathcal{L}}{%
dF^{2}}\right\vert _{F^{2}=-2E^{2}}\right) =0\,,  \label{Et}
\end{equation}%
whose solution introduces $q$ as an integration constant related to the
electric charge,%
\begin{equation}
E\,\left. \frac{d\mathcal{L}}{dF^{2}}\right\vert _{F^{2}=-2E^{2}}=-\frac{q}{%
r^{D-2}}\,.  \label{r(E)}
\end{equation}%
The electric field decouples from the metric and one can determine $E$ just
from this algebraic equation. Therefore, the electric potential at the
distance $r$ measured with respect to radial infinity is calculated from%
\begin{equation}
\phi (r)=-\int\limits_{\infty }^{r}dv\,E(v)\,.  \label{potential}
\end{equation}%
For the analysis of thermodynamic properties of charged black holes, the
conjugate variable to the electric charge $Q$ is the difference in electric
potential between infinity and the event horizon $r_{+}$, that is, $\Phi
=\phi (\infty )-\phi (r_{+})$.

The first integral of the equation of motion for the metric (\ref{grav diff
Eq.}) is%
\begin{equation}
\left( f^{2}-k\right) \left( 1-\alpha \left( D-3\right) \left( D-4\right) \,%
\frac{f^{2}-k}{r^{2}}\right) =\frac{r^{2}}{\ell ^{2}}-\frac{\mu }{r^{D-3}}+%
\frac{16\pi G\,\mathcal{T}(q,r)}{\left( D-2\right) r^{D-3}}\,,
\label{f quadratic}
\end{equation}%
where $\mu $ is related to the black hole mass, whereas the electromagnetic
flux through the surface $r=const$ is given by the function%
\begin{eqnarray}
\mathcal{T}(q,r) &=&\int\limits_{\infty }^{r}dv\,v^{D-2}\,T_{r}^{r}
\nonumber \\
&=&\int\limits_{\infty }^{r}dv\,\left( \rule{0in}{0.18in}v^{D-2}\mathcal{L}%
-4qE\right) \,,  \label{I(q,r)}
\end{eqnarray}%
for an arbitrary NED Lagrangian. In the second line, the charge $q$ appears
due to the Gauss law (\ref{r(E)}).

As mentioned above, the theory possesses two branches, what is manifested as
two solutions in the metric function $f^{2}$ of Eq.(\ref{f quadratic}) for
EGB gravity coupled to NED,
\begin{equation}
f_{\pm }^{2}(r)=k+\frac{r^{2}}{2\alpha \left( D-3\right) \left( D-4\right) }%
\left[ 1\pm \sqrt{1-4\alpha \left( D-3\right) \left( D-4\right) \left( \frac{%
1}{\ell ^{2}}-\frac{\mu }{r^{D-1}}+\frac{16\pi G\,\mathcal{T}(q,r)}{\left(
D-2\right) r^{D-1}}\right) }\right] \,.  \label{2 metric functions}
\end{equation}

The branch with positive sign $f_{+}^{2}(r)$ has an ill-defined limit $%
\alpha \rightarrow 0$, and it does not recover the solutions of EH AdS
gravity. Furthermore, perturbations around the corresponding vacuum state
are proved to have negative mass \cite{Boulware-Deser}. Henceforth, we will
focus on the stable and ghost-free branch $f^{2}(r)\equiv f_{-}^{2}(r)$.
Asymptotically, this branch and its radial derivative behave as\textbf{\ }%
\begin{eqnarray}
f^{2} &=&k+\frac{r^{2}}{\ell _{e\!f\!f}^{2}}-\frac{\mu }{1-\frac{2\alpha }{%
\ell _{e\!f\!f}^{2}}\,\left( D-3\right) \left( D-4\right) }\frac{1}{r^{D-3}}+%
\mathcal{O}\left( \frac{1}{r^{2D-6}}\right) \,,  \label{f_asympt} \\
(f^{2})^{\prime } &=&\frac{2r}{\ell _{e\!f\!f}^{2}}+\frac{\left( D-3\right)
\mu }{1-\frac{2\alpha }{\ell _{e\!f\!f}^{2}}\,\left( D-3\right) \left(
D-4\right) }\frac{1}{r^{D-2}}+\mathcal{O}\left( \frac{1}{r^{2D-5}}\right) \,.
\label{df_asympt}
\end{eqnarray}

The fact that there are no additional contributions to the energy of charged
black holes in any physically sensible NED theory is a consequence of the
metric fall-off, which is the same as in Reissner-Nordstrom case.

The procedure outlined above can be regarded as an algorithm to construct
explicit solutions to various NED theories: conformally invariant
electrodynamics, Born-Infeld, Logarithmic electrodynamics, etc., as it was
pointed out in Ref. \cite{Miskovic-OleaNEDcharges}.

\section{Iyer-Wald charges and black hole entropy}

At this point, we proceed to evaluate the Euclidean continuation of the
action for EGB gravity coupled to NED in Eqs.(\ref{EGB AdS}) and (\ref{NED}).

As a first step, without loss of generality, we take Gauss-normal
coordinates to foliate the spacetime along a radial direction
\begin{equation}
ds^{2}=g_{\mu \nu }\,dx^{\mu }dx^{\nu }=N^{2}\left( r\right)
\,dr^{2}+h_{ij}(r,x)\,dx^{i}dx^{j}\,.  \label{Gaussian}
\end{equation}%
This frame is obtained as a particular gauge-fixing ($N^{i}=0$) of a general
ADM form of the metric. We assume the manifold to have a boundary $\partial
\mathcal{M}$, located at radial infinity, endowed with an induced metric $%
h_{ij}$.

The identities for the totally antisymmetric product of Kronecker deltas
listed in Appendix \ref{Delta} enable us to write down the pure gravity part
of the bulk action as
\begin{eqnarray}
I_{grav} &=&\frac{1}{16\pi G\left( D-2\right) !}\int\limits_{\mathcal{M}%
}d^{D}x\,\sqrt{-g}\,\delta _{\lbrack \nu _{1}\cdots \nu _{D}]}^{[\mu
_{1}\cdots \mu _{D}]}\,\left( \frac{1}{2}\,R_{\mu _{1}\mu _{2}}^{\nu _{1}\nu
_{2}}\,\delta _{\mu _{3}}^{\nu _{3}}\delta _{\mu _{4}}^{\nu _{4}}\right.
\nonumber \\
&&\qquad +\left. \frac{D-2}{D\,\ell ^{2}}\,\delta _{\mu _{1}}^{\nu
_{1}}\delta _{\mu _{2}}^{\nu _{2}}\delta _{\mu _{3}}^{\nu _{3}}\delta _{\mu
_{4}}^{\nu _{4}}+\frac{\alpha \left( D-2\right) \left( D-3\right) }{4}%
\,R_{\mu _{1}\mu _{2}}^{\nu _{1}\nu _{2}}R_{\mu _{3}\mu _{4}}^{\nu _{3}\nu
_{4}}\right) \,\delta _{\mu _{5}}^{\nu _{5}}\cdots \delta _{\mu _{D}}^{\nu
_{D}}\,.  \label{EH-GB}
\end{eqnarray}

In the foliation (\ref{Gaussian}), the indices split as $\mu =\left(
r,i\right) $, where Latin letters denote boundary components. Plugged in the
total action Eq.(\ref{bulk action}), this evaluation produces
\begin{eqnarray}
I_{0} &=&\frac{1}{16\pi G\left( D-3\right) \left( D-4\right) }\int\limits_{%
\mathcal{M}}d^{D-1}x\,dr\,\sqrt{-h}\,N\,\delta _{\lbrack i_{1}\cdots
i_{4}]}^{[j_{1}\cdots j_{4}]}\left[ \left( \frac{1}{2}%
\,R_{j_{1}j_{2}}^{i_{1}i_{2}}\,+\frac{1}{\ell ^{2}}\,\delta
_{j_{1}}^{i_{1}}\delta _{j_{2}}^{i_{2}}+\frac{2}{D-2}\,R_{ri_{1}}^{ri_{1}}%
\delta _{j_{2}}^{i_{2}}\right) \delta _{j_{3}}^{i_{3}}\delta
_{j_{4}}^{i_{4}}\right.  \nonumber \\
&&+\left. \alpha \left( D-3\right) \left( \frac{D-4}{4}%
\,R_{j_{1}j_{2}}^{i_{1}i_{2}}+2R_{rj_{1}}^{ri_{1}}\,\delta
_{j_{2}}^{i_{2}}\right) R_{j_{3}j_{4}}^{i_{3}i_{4}}\right] +\int\limits_{%
\mathcal{M}}d^{D-1}x\,dr\,\sqrt{-h}\,N\mathcal{L}(F^{2})\,,
\end{eqnarray}%
where we have used the fact that $\delta _{\lbrack r\,i_{1}\cdots
i_{4}]}^{[r\,j_{1}\cdots j_{4}]}=\delta _{\lbrack i_{1}\cdots
i_{4}]}^{[j_{1}\cdots j_{4}]}$. One can recognize the component $\mathcal{E}%
_{r}^{r}$ of the equations of motion in the form (\ref{Emu_nu}) (see
Appendix \ref{Delta}) from the first two terms in the first line and the
first term in the second one. In doing so, the action in radial normal frame
reads
\begin{eqnarray}
I_{0} &=&\frac{1}{8\pi G\left( D-2\right) \left( D-3\right) }\int\limits_{%
\mathcal{M}}d^{D-1}x\,dr\,\sqrt{-h}\,N\,\delta _{\lbrack i_{1}\cdots
i_{3}]}^{[j_{1}\cdots j_{3}]}\,R_{ri_{1}}^{ri_{1}}\left( \delta
_{j_{2}}^{i_{2}}\delta _{j_{3}}^{i_{3}}+\alpha \left( D-2\right) \left(
D-3\right) R_{j_{2}j_{3}}^{i_{2}i_{3}}\right)  \nonumber \\
&&\qquad \qquad \qquad -\int\limits_{\mathcal{M}}d^{D-1}x\,dr\,\sqrt{-h}%
\,N\,\left( \frac{1}{8\pi G}\,\mathcal{E}_{r}^{r}+T_{r}^{r}-\mathcal{L}%
(F^{2})\right) \,.
\end{eqnarray}%
We consider a spacetime given by the Euclidean continuation of the static
black hole metric (\ref{BH}), where the Euclidean time $\tau =it$ appears as
identified in a period $\beta $, which is the inverse of the Hawking
temperature%
\begin{equation}
T\equiv \beta ^{-1}=\frac{1}{4\pi }\,\left. \frac{df^{2}(r)}{dr}\right\vert
_{r=r_{+}}\,.  \label{HawkingT}
\end{equation}%
As the horizon is reduced to a single point, the Euclidean period in Eq.(\ref%
{HawkingT}) smooths out the origin of the radial coordinate $r=r_{+}$,
avoiding the presence of a cone-like singularity. We stress the fact that no
boundary is introduced at the horizon, such that no extra surface terms are
needed there in order to reproduce the correct black hole thermodynamics.

More explicitly, the Hawking temperature is given by%
\begin{equation}
T=\frac{1}{4\pi r_{+}}\frac{\left( D-3\right) k+\frac{\left( D-1\right)
\,r_{+}^{2}}{\ell ^{2}}+\alpha \,\left( D-3\right) \left( D-4\right) \left(
D-5\right) \,\frac{k^{2}}{r_{+}^{2}}+\frac{16\pi G\,r_{+}^{2}}{D-2}%
\,T_{r}^{r}(q,r_{+})}{1+2\alpha \,\left( D-3\right) \left( D-4\right) \,%
\frac{k}{r_{+}^{2}}}\,,  \label{T}
\end{equation}%
where the electromagnetic field contributes to the black hole temperature
trough the tensor $T_{r}^{r}(q,r_{+})=\mathcal{L}+4E^{2}\frac{d\mathcal{L}}{%
dF^{2}}$ evaluated at $F^{2}=-2E^{2}$ and for $r=r_{+}$.

The Wick rotation implies $I^{E}=-iI$ \ for the Euclidean action, that can
be expressed in terms of the black hole metric function $f^{2}$ and its
derivatives as%
\begin{eqnarray}
I_{0}^{E} &=&\frac{1}{16\pi G}\int\limits_{0}^{\beta }d\tau
\int\limits_{\Gamma _{D-2}}\sqrt{\gamma }\,d^{D-2}\varphi
\int\limits_{r_{+}}^{\infty }dr\,r^{D-2}\,\left[ \rule{0in}{0.27in}%
(f^{2})^{\prime \prime }+\left( D-2\right) \,\frac{(f^{2})^{\prime }}{r}%
+\right.   \nonumber \\
&&\left. +\frac{2\alpha \left( D-2\right) \left( D-3\right) }{r^{2}}\,\left(
(f^{2})^{\prime \prime }\left( k-f^{2}\right) -(f^{2})^{\prime 2}+\left(
D-4\right) \,\frac{(f^{2})^{\prime }\left( k-f^{2}\right) }{r}\right) \right]
-  \nonumber \\
&&+4\int\limits_{0}^{\beta }d\tau \int\limits_{\Gamma _{D-2}}\sqrt{\gamma }%
\,d^{D-2}\varphi \int\limits_{r_{+}}^{\infty }dr\,r^{D-2}\,E^{2}\left. \frac{%
d\mathcal{L}}{dF^{2}}\right\vert _{F^{2}=-2E^{2}}\,,
\end{eqnarray}%
where we have substituted the expressions for the Riemann tensor in Appendix %
\ref{TBH metric} and the electromagnetic stress tensor (\ref{T components}).
Upon a trivial integration on the boundary coordinates, all the integrand
can be written as a total derivative, and the bulk Euclidean action becomes
\begin{eqnarray}
I_{0}^{E} &=&\frac{\beta \,\text{Vol}(\Gamma _{D-2})}{16\pi G}\,\left. \left[
(f^{2})^{\prime }\left( \rule{0in}{0.2in}r^{D-2}+2\alpha \left( D-2\right)
\left( D-3\right) r^{D-4}\left( k-f^{2}\right) \right) \right] \right\vert
_{r_{+}}^{\infty }  \nonumber \\
&&-\beta \,\text{Vol}(\Gamma _{D-2})\left. \left( 4r^{D-2}\phi \,E\,\frac{d%
\mathcal{L}}{dF^{2}}\right) \right\vert _{r_{+}}^{\infty }\,,
\label{Euclid_bulk}
\end{eqnarray}%
where, in the last line, the equation of motion for NED field (\ref{Et}) was
used.

By definition of the Euclidean period $\beta $, the first line evaluated at
the horizon $r_{+}$ produces $-S$, where $S$ is the standard value of black
hole entropy in EGB gravity \cite{Cai,Cvetic-Nojiri-Odintsov}
\begin{equation}
S=\frac{\text{Vol}(\Gamma _{D-2})\,r_{+}^{D-2}}{4G}\,\left( 1+\frac{2k\alpha
}{r_{+}^{2}}\,\left( D-2\right) \left( D-3\right) \right) \,,  \label{S EGB}
\end{equation}%
whereas the second line is $\beta Q\Phi $, where
\begin{equation}
Q=4\text{Vol}(\Gamma _{D-2})\,q\,.  \label{Qdef}
\end{equation}%
Notice that, what is physically relevant for the thermodynamic description
of NED action, is the potential difference between the horizon and infinity.
Therefore, we assume $\phi (\infty )=0$ without loss of generality.

Thus, the Euclidean continuation of the bulk action for EGB AdS gravity
coupled to NED is written as%
\begin{equation}
I_{0}^{E}=\beta Q\Phi -S+\frac{\beta \,\text{Vol}(\Gamma _{D-2})}{16\pi G}%
\,\lim_{r\rightarrow \infty }\left[ (f^{2})^{\prime }\left( \rule{0in}{0.2in}%
r^{D-2}+2\alpha \left( D-2\right) \left( D-3\right) r^{D-4}\left(
k-f^{2}\right) \right) \right] ,
\end{equation}%
what, in terms of the black hole mass \cite{Kofinas-Olea-GB,Deser-Tekin,
Padilla, Gravanis}%
\begin{equation}
M=(D-2)\,\frac{\text{Vol}(\Gamma _{D-2})\,\mu }{16\pi G}\,,  \label{BH Mass}
\end{equation}%
can be recast as%
\begin{eqnarray}
I_{0}^{E} &=&\beta Q\Phi -S+\beta \,M\,\frac{D-2}{D-3}\frac{\ell
_{e\!f\!f}^{2}-2\alpha \,(D-2)(D-5)}{\ell _{e\!f\!f}^{2}-2\alpha (D-3)(D-4)}+
\nonumber \\
&&+\frac{\beta \,\text{Vol}(\Gamma _{D-2})}{16\pi G}\,\lim_{r\rightarrow
\infty }\left[ \frac{r^{D-3}}{\ell _{e\!f\!f}^{2}}\left( \rule{0in}{0.2in}%
1+2\alpha \left( D-2\right) \left( D-3\right) r^{D-4}\left( k-f^{2}\right)
\right) \right] .  \label{IE divergent}
\end{eqnarray}%
The unusual factor multiplying $\beta M$ indicates that a
background-subtraction method would not necessarily give rise to a correct
QSR. Indeed, subtracting the value of $I_{0}^{E}$ evaluated for AdS vacuum
(with the corresponding AdS radius $\ell _{e\!f\!f}$) gets rid of the
divergences at $r=\infty $, but does not reproduce the mass in the
asymptotic region. The quantity that appears at radial infinity is the
analog of Komar formula for EGB gravity \cite{Kastor}.

It is possible to understand the above result in the light of Iyer-Wald
definition of conserved quantities in gravity theories \cite{Iyer-Wald}.
This time-honored procedure interprets the black hole entropy as the Noether
charge $\tilde{Q}\left[ \xi \right] $ for a Killing vector $\xi =\xi ^{\mu
}\partial _{\mu }$, evaluated at the horizon. This quantity, however, is
derived exclusively from the bulk Lagrangian of the gravity theory (which in
our case is denoted by $\mathcal{L}_{grav}$), without additional boundary
terms. More concretely, the charge $\tilde{Q}\left[ \xi \right] $ is written
in terms of the derivative of the bulk Lagrangian respect to the Lorentz
curvature two-form. In tensorial notation, we can work out an alternative
form if the Lagrangian density is re-written as $\mathcal{L}_{grav}=\delta
_{\lbrack \nu _{1}\cdots \nu _{D}]}^{[\mu _{1}\cdots \mu _{D}]}L_{\mu
_{1}\cdots \mu _{D}}^{\nu _{1}\cdots \nu _{D}}$. In this way, the charge is
given by the fully-covariant formula in terms of derivatives of the Riemann
tensor
\begin{equation}
\tilde{Q}\left[ \xi \right] =\int\limits_{\partial \mathcal{M}_{r}\cap \Xi
_{t}}d^{D-2}x\,\sqrt{-g}\,\hat{n}_{\mu }\,\hat{u}_{\nu }\,\xi ^{\lambda
}\,\delta _{\lbrack \nu _{1}\cdots \nu _{D}]}^{[\mu _{1}\cdots \mu
_{D}]}\,g^{\alpha \sigma }\,\Gamma _{\lambda \sigma }^{\beta }\frac{\delta
L_{\mu _{1}\cdots \mu _{D}}^{\nu _{1}\cdots \nu _{D}}}{\delta R_{\mu \nu
}^{\alpha \beta }}\,,  \label{I-W tensor}
\end{equation}%
which must be evaluated at the intersection of a $r=const.$ boundary $%
\partial \mathcal{M}_{r}$ with a constant-time slice $\Xi _{t}$ \cite%
{Iyer-Wald,Clunan-Ross-Smith,Paranjape-Sarkar-Padmanabhan}. The normal
vectors $\hat{n}_{\mu }\,$\ and $\hat{u}_{\mu }$ generate the corresponding
foliations which describe $\partial \mathcal{M}_{r}$ and $\Xi _{t}$,
respectively \cite{footnote3}.

The form of the action (\ref{EH-GB}) for EGB gravity is particularly useful
to compute the above expression, what produces%
\begin{eqnarray}
\tilde{Q}\left[ \xi \right]  &=&-\frac{1}{16\pi G\left( D-2\right) \left(
D-3\right) }\int\limits_{\partial \mathcal{M}_{r}\mathcal{\cap }\Xi
_{t}}d^{D-2}x\,\sqrt{-g}\,\hat{n}_{\mu _{1}}\,\hat{u}_{\mu _{2}}\,\xi
^{\lambda }\,\delta _{\lbrack \nu _{1}\nu _{2}\nu _{3}\nu _{4}]}^{[\mu
_{1}\mu _{2}\mu _{3}\mu _{4}]}\,g^{\nu _{1}\alpha }\,\Gamma _{\lambda \alpha
}^{\nu _{2}}\times   \nonumber \\
&&\qquad \times \left( \delta _{\lbrack \mu _{3}\mu _{4}]}^{[\nu _{3}\nu
_{4}]}+2\left( D-2\right) \left( D-3\right) \alpha R_{\mu _{3}\mu _{4}}^{\nu
_{3}\nu _{4}}\right) \,.  \label{I-W EGB}
\end{eqnarray}%
In the Gauss-normal frame (\ref{Gaussian}), the normal derivative of the
induced metric $h_{ij}$ describes the extrinsic properties of the boundary,%
\begin{equation}
K_{ij}=-\frac{1}{2N}\,h_{ij}^{\prime }\,,
\end{equation}%
what defines the extrinsic curvature. Then the Christoffel symbol can be
expressed in terms of $K_{ij}$ (see Appendix \ref{Gauss-normalApp}), such
that the Iyer-Wald charge is%
\begin{eqnarray}
\tilde{Q}\left[ \xi \right]  &=&\frac{1}{16\pi G\left( D-2\right) \left(
D-3\right) }\,\int\limits_{\partial \mathcal{M}_{r}\mathcal{\cap }\Xi
_{t}}d^{D-2}x\,\sqrt{-h}\,\hat{u}_{j}\,\xi ^{i}\,\delta _{\lbrack
i_{1}\cdots i_{D-1}]}^{[jj_{2}\cdots j_{D-1}]}\,K_{i}^{i_{1}}\times
\nonumber \\
&&\qquad \times \left( \delta _{\lbrack j_{2}j_{3}]}^{[i_{2}i_{3}]}+2\alpha
\left( D-2\right) \left( D-3\right) \,R_{j_{2}j_{3}}^{i_{2}i_{3}}\right) ,
\label{I-W EGB boundary}
\end{eqnarray}%
where$\ \sqrt{-g}=N\sqrt{-h}$. It is not difficult to show that, for the
black hole metric (\ref{BH}), the Noether charge (\ref{I-W EGB boundary})
leads to
\begin{equation}
\tilde{Q}\left[ \partial _{t}\right] =\frac{\text{Vol}(\Gamma _{D-2})}{16\pi
G}\,(f^{2})^{\prime }\left( \rule{0in}{0.2in}r^{D-2}+2\alpha \left(
D-2\right) \left( D-3\right) r^{D-4}\left( k-f^{2}\right) \right) \,,
\end{equation}%
for an arbitrary surface $\Gamma _{D-2}$ of radius $r$. This quantity
clearly reproduces the value of the entropy (\ref{S EGB}) when evaluated at
the horizon $r=r_{+}$, and a finite contribution (which cannot be identified
with the mass) plus a divergent term at infinity, which are the same terms
present in Eq.(\ref{IE divergent}). The fact that $I_{0}^{E}$ can be written
down as
\begin{equation}
I_{0}^{E}=\beta Q\Phi +\beta \left. \tilde{Q}\left[ \partial _{t}\right]
\right\vert _{r_{+}}^{\infty }\,,
\end{equation}%
means that the problem of finiteness of the Euclidean action is connected to
the regularization of the Noether charges in the asymptotic region.

It is possible to correct the Iyer-Wald formula, such that it provides the
correct mass and angular momentum for black holes, through the construction
of a Hamiltonian $H\left[ \xi \right] $ which describes the dynamics
generated by the vector field $\xi ^{\mu }$. Generally speaking, when one
varies the action respect to the fields $\phi $, one identifies the
equations of motion plus a surface term%
\begin{equation}
\delta I=\int\limits_{\mathcal{M}}\left( EOM\right) \,\delta \phi
+\int\limits_{\partial \mathcal{M}}\Theta (\phi ,\delta \phi )\,.
\label{deltaI}
\end{equation}%
Then, the Hamiltonian is related to the Iyer-Wald charge at radial infinity
$\tilde{Q}_{\infty }\left[ \xi \right] =\int_{\Sigma _{\infty }}
\tilde{\mathcal{Q}}(\xi )$ and surface term in eq.(\ref{deltaI}) by
\begin{equation}
\delta H\left[ \xi \right] =\delta \int\limits_{\Sigma _{\infty }}
\tilde{\mathcal{Q}}(\xi )-\int\limits_{\Sigma _{\infty }}\xi \cdot \Theta (\phi
,\delta \phi )\,,
\end{equation}%
where $\Sigma _{\infty }=\partial \mathcal{M}\cap \Xi _{t}$. The Hamiltonian
exists if there is a $(D-1)$-form $\mathcal{B}$ such that the second term is
a total variation,%
\begin{equation}
\int\limits_{\Sigma _{\infty }}\xi \cdot \Theta (\phi ,\delta \phi )=\delta
\int\limits_{\Sigma _{\infty }}\xi \cdot \mathcal{B}(\phi )\,.
\label{integrability}
\end{equation}%
Whenever this is possible, one can write down the Wald Hamiltonian as \cite%
{Iyer-Wald}%
\begin{equation}
H\left[ \xi \right] =\int\limits_{\Sigma _{\infty }}\left(
\tilde{\mathcal{Q}}(\xi )-\xi \cdot \mathcal{B}\right) \,.  \label{Wald_H}
\end{equation}

However, the procedure outlined above in general breaks covariance at the
boundary. As a result, the correction $\mathcal{B}$ to the charge has a
non-covariant form, and has to be built on a case-by-case basis for
different solutions. Moreover, if exists, $\mathcal{B}$ cannot be used as an
appropriate boundary term to render the Euclidean action finite.

In what follows, we employ covariant counterterms given as polynomials in
the extrinsic and intrinsic curvatures to regularize the Euclidean action
and to obtain the QSR given by Eq.(\ref{QSR1}).

\section{Kounterterm regularization and Quantum Statistical Relation}

The regularization of gravitational action prescribed by AdS/CFT
correspondence leads to the addition of covariant functionals of the
boundary metric and curvature, known as counterterm series. In a dual
quantum field theory, this counterterm series corresponds to a standard UV
divergence removal by adding finite polynomials in the fields.

However, in EGB-AdS gravity, holographic renormalization becomes too
involved to provide a general answer to this problem.

Here, we use an alternative regularization procedure, where the counterterms
depend explicitly on the extrinsic curvature. The choice of such terms in
AdS gravity is justified by the asymptotic expansions of the fields. Indeed,
the extrinsic curvature and the boundary metric are proportional at the
leading order. In that way, the boundary term $B_{D-1}$ is given as a unique
geometrical structure depending only on the dimension. This approach
circumvents the technicalities of holographic methods for higher-curvature
theories of the Lovelock class \cite{Kofinas-Olea}.

We will work with gravity-NED action in $D$ dimensions supplemented by a
boundary term
\begin{equation}
I=I_{0}+c_{D-1}\int\limits_{\partial \mathcal{M}}d^{D-1}x\,B_{D-1}\,,
\label{I}
\end{equation}%
where the coupling $c_{D-1}$ is fixed demanding a well-defined action
principle \cite{Kofinas-Olea-GB,Miskovic-OleaNEDcharges}. The explicit form
of the Kounterterm series as a polynomial of the extrinsic and intrinsic
curvatures for EH AdS gravity was introduced in Refs.\cite%
{Olea-K,Olea-KerrBH}.

For the current discussion, we consider the grand canonical ensemble, where
the temperature $T$ and the electric potential $\Phi $ are held fixed at the
horizon. The Gibbs free energy
\begin{equation}
G(T,\Phi )=U-TS+\Phi Q\,,  \label{Gdef}
\end{equation}%
that satisfies the differential equation
\begin{equation}
dG=-S\,dT+Qd\Phi \,,  \label{dG}
\end{equation}%
is given in terms of the Euclidean action as
\begin{equation}
G=T\,I^{E}\,.  \label{Gibbs}
\end{equation}

When the partition function in semiclassical approximation is written in
terms of Gibbs energy as
\begin{equation}
Z=e^{-G/T}\,.
\end{equation}

In what follows, we employ Kounterterms to regulate the value of the
Euclidean action for EGB-AdS coupled to NED and to obtain the QSR (\ref{QSR1}%
) for charged black hole solutions.

\subsection{Even dimensions}

In a manifold $\mathcal{M}$ without boundary in even dimensions $D=2n$, the
integration of the Euler topological invariant produces the Euler
characteristic, $\chi (\mathcal{M})$. If a boundary $\partial \mathcal{M}$
is introduced, there appears a correction to $\chi (\mathcal{M})$ given by
the $n$-th Chern form%
\begin{equation}
B_{2n-1}=2n\sqrt{-h}\int\limits_{0}^{1}dt\,\delta _{\lbrack i_{1}\cdots
i_{2n-1}]}^{[j_{1}\cdots j_{2n-1}]}\,K_{j_{1}}^{i_{1}}\left( \frac{1}{2}\,%
\mathcal{R}%
_{j_{2}j_{3}}^{i_{2}i_{3}}-t^{2}K_{j_{2}}^{i_{2}}K_{j_{3}}^{i_{3}}\right)
\cdots \left( \frac{1}{2}\,\mathcal{R}%
_{j_{2n-2}j_{2n-1}}^{i_{2n-2}i_{2n-1}}-t^{2}K_{j_{2n-2}}^{i_{2n-2}}K_{j_{2n-1}}^{i_{2n-1}}\right) \,.
\label{B-even}
\end{equation}%
Here, $\mathcal{R}_{kl}^{ij}(h)$ is the intrinsic curvature of the boundary,
related to the spacetime Riemann tensor by $R_{kl}^{ij}=\mathcal{R}%
_{kl}^{ij}-K_{k}^{i}K_{l}^{j}+K_{l}^{i}K_{k}^{j}$ (see Appendix \ref%
{Gauss-normalApp}). In asymptotically AdS spacetimes, the addition of the
above boundary term to the gravity action defines an extrinsic
regularization scheme in even dimensions. In EGB AdS gravity, it has been
shown in Ref.\cite{Kofinas-Olea-GB} that the constant $c_{2n-1}$ in front of
the boundary term $B_{2n-1}$ has to be fixed in terms of the effective AdS
radius as
\begin{equation}
c_{2n-1}=-\frac{1}{16\pi G}\frac{(-\ell _{e\!f\!f}^{2})^{n-1}}{n\left(
2n-2\right) !}\left( 1-\frac{2\alpha }{\ell _{e\!f\!f}^{2}}\,\left(
2n-2\right) \left( 2n-3\right) \right) \,,  \label{c2n-1}
\end{equation}%
in order to cancel the divergences in the Euclidean action. Notice that the
integration on the continuous parameter $t$ generates the coefficients if
one wants to express the boundary term as a polynomial. This compact way of
writing $B_{2n-1}$ is not a mere formality but it reflects the relation to
topological invariants and provides a useful tool for explicit evaluations,
as well.

The boundary term (\ref{B-even}) evaluated on the Euclidean continuation of
the black hole metric (\ref{BH}) becomes%
\begin{eqnarray}
\int\limits_{\partial \mathcal{M}}d^{2n-1}x\,B_{2n-1}^{E}
&=&-2n\lim_{r\rightarrow \infty }\int\limits_{0}^{\beta }d\tau
\int\limits_{\Gamma _{D-2}}\sqrt{\gamma }\,d^{D-2}\varphi
\,r^{2n-2}f\int\limits_{0}^{1}dt\,\delta _{\lbrack n_{1}\cdots
n_{2n-2}]}^{[m_{1}\cdots m_{2n-2}]}  \nonumber \\
&&\times K_{\tau }^{\tau }\left( \frac{1}{2}\,\mathcal{R}%
_{m_{1}m_{2}}^{n_{1}n_{2}}-\left( 2n-1\right)
t^{2}K_{m_{1}}^{n_{1}}K_{m_{2}}^{n_{2}}\right) \times   \nonumber \\
&&\left( \frac{1}{2}\,\mathcal{R}%
_{m_{3}m_{4}}^{n_{3}n_{4}}-t^{2}K_{m_{3}}^{n_{3}}K_{m_{4}}^{n_{4}}\right)
\cdots \left( \frac{1}{2}\,\mathcal{R}%
_{m_{2n-3}m_{2n-2}}^{n_{2n-3}n_{2n-2}}-t^{2}K_{m_{2n-3}}^{n_{2n-3}}K_{m_{2n-2}}^{n_{2n-2}}\right) .
\end{eqnarray}%
Substituting the components of the extrinsic curvature (\ref{KBH}) and
intrinsic curvature (\ref{calR}) (see Appendix \ref{TBH metric}) and
performing the integral
\begin{equation}
\int\limits_{0}^{1}dt\,\left[ k-\left( 2n-1\right) t^{2}f^{2}\right] \left(
k-t^{2}f^{2}\right) ^{n-2}=\left( k-f^{2}\right) ^{n-1}\,,
\end{equation}%
the boundary term takes the form
\begin{equation}
c_{2n-1}\int\limits_{\partial \mathcal{M}}d^{2n-1}x\,B_{2n-1}^{E}=\frac{%
\beta \,\text{Vol}(\Gamma _{D-2})\,\ell _{e\!f\!f}^{2n-2}}{16\pi G}\,\left(
1-\frac{2\alpha }{\ell _{e\!f\!f}^{2}}\,\left( 2n-2\right) \left(
2n-3\right) \right) \left. (f^{2})^{\prime }\left( f^{2}-k\right)
^{n-1}\right\vert ^{r=\infty }\,.
\end{equation}%
In consequence, the total Euclidean action
\begin{equation}
I_{2n}^{E}=I_{0}^{E}+c_{2n-1}\int\limits_{\partial \mathcal{M}%
}d^{2n-1}x\,B_{2n-1}^{E}\,,
\end{equation}%
can be written as%
\begin{eqnarray}
I_{2n}^{E} &=&\frac{\beta \,\text{Vol}(\Gamma _{D-2})}{16\pi G}\left\{ \,%
\rule{0in}{0.26in}\left. (f^{2})^{\prime }\left( \rule{0in}{0.2in}%
r^{D-2}+2\alpha \left( 2n-2\right) \left( 2n-3\right) r^{D-4}\left(
k-f^{2}\right) \right) \right\vert _{r_{+}}^{\infty }\right.   \nonumber \\
&&\left. -\ell _{e\!f\!f}^{2n-2}\,\left( 1-\frac{2\alpha }{\ell
_{e\!f\!f}^{2}}\,\left( 2n-2\right) \left( 2n-3\right) \right) \left. \left[
(f^{2})^{\prime }\left( f^{2}-k\right) ^{n-1}\right] \right\vert ^{r=\infty
}\right\} +\beta Q\Phi \,.  \label{IE2n}
\end{eqnarray}%
The contribution at infinity from the bulk action combines with the one from
the boundary to produce $\beta $ times the Noether mass $M$
\begin{eqnarray}
M &=&\frac{\text{Vol}(\Gamma _{2n-2})}{16\pi G}\,\lim_{r\rightarrow \infty
}\,r^{2n-2}(f^{2})^{\prime }\left[ 1-2\alpha \left( 2n-2\right) \left(
2n-3\right) \,\frac{f^{2}-k}{r^{2}}-\right.   \nonumber \\
&&\hspace{-0.5cm}\left. -\left( 1-\frac{2\alpha \left( 2n-2\right) \left(
2n-3\right) }{\ell _{e\!f\!f}^{2}}\right) \ell _{e\!f\!f}^{2n-2}\left( \frac{%
f^{2}-k}{r^{2}}\right) ^{n-1}\right] \,,
\end{eqnarray}%
which agrees with the formula (\ref{BH Mass}) when expressed in terms of $%
\mu $ \cite{Miskovic-OleaNEDcharges}. In other words, because Kounterterm
series leads to the cancellation of divergences in the asymptotic charges,
the finiteness of the Euclidean action is ensured for any static black hole
and satisfies the QSR%
\begin{equation}
G(T,\Phi )=T\,I_{2n}^{E}=U-T\,S+Q\Phi \,,
\end{equation}%
where the EGB black hole entropy is given by Eq.(\ref{S EGB}).

\subsection{Odd dimensions}

In $D=2n+1$ dimensions, Kounterterm regularization provides the explicit
form of the boundary terms which makes the Euclidean action finite in EH AdS
gravity \cite{Olea-K}. The universality of the Kounterterm series ensures
that the action will be also finite in any Lovelock gravity with AdS
branches. That means that the information on a particular theory is
incorporated in the regularization scheme through effective AdS radius $\ell
_{e\!f\!f}$ and the coupling constant $c_{2n}$. Thus, in general, the
Kounterterms series is given in terms of the parametric integrations%
\begin{equation}
B_{2n}=\frac{n}{2^{n-2}}\,\sqrt{-h}\int\limits_{0}^{1}dt\int%
\limits_{0}^{t}ds\,\delta _{\lbrack i_{1}\cdots i_{2n}]}^{[j_{1}\cdots
j_{2n}]}\,K_{j_{1}}^{i_{1}}\delta _{j_{2}}^{i_{2}}\,\mathcal{F}%
_{j_{3}j_{4}}^{i_{3}i_{4}}(t,s)\cdots \mathcal{F}%
_{j_{2n-1}j_{2n}}^{i_{2n-1}i_{2n}}(t,s)\,,  \label{B-odd}
\end{equation}%
where we introduce an auxiliary quantity with the symmetries of the Riemann
tensor%
\begin{equation}
\mathcal{F}_{kl}^{ij}(t,s)=\mathcal{R}_{kl}^{ij}-t^{2}\left(
K_{k}^{i}K_{l}^{j}-K_{l}^{i}K_{k}^{j}\right) +\frac{s^{2}}{\ell
_{e\!f\!f}^{2}}\,\delta _{\lbrack kl]}^{[ij]}\,,  \label{Fst}
\end{equation}%
such that in EGB AdS theory the coupling reads \cite{Kofinas-Olea-GB}
\begin{eqnarray}
c_{2n} &=&-\frac{1}{16\pi G}\frac{(-\ell _{e\!f\!f}^{2})^{n-1}}{n\left(
2n-1\right) !}\left( 1-\frac{2\alpha \left( 2n-1\right) \left( 2n-2\right) }{%
\ell _{e\!f\!f}^{2}}\right) \left[ \int\limits_{0}^{1}dt\,\left(
1-t^{2}\right) ^{n-1}\right] ^{-1}  \nonumber \\
&=&-\frac{1}{16\pi G}\frac{2(-\ell _{e\!f\!f}^{2})^{n-1}}{n\left(
2n-1\right) !\beta (n,\frac{1}{2})}\left( 1-\frac{2\alpha \left( 2n-1\right)
\left( 2n-2\right) }{\ell _{e\!f\!f}^{2}}\right) \,,  \label{c2n}
\end{eqnarray}%
where $\beta (n,\frac{1}{2})=\frac{2^{2n-1}\left( n-1\right) !^{2}}{\left(
2n-1\right) !}$ is the Beta function for those arguments.

In the Euclidean sector, the boundary term in the black hole ansatz (\ref{BH}%
) produces
\begin{eqnarray}
\int\limits_{\partial \mathcal{M}}d^{2n}x\,B_{2n}^{E} &=&-\frac{n}{2^{n-3}}%
\,\lim_{r\rightarrow \infty }\int\limits_{0}^{\beta }d\tau
\int\limits_{\Gamma _{D-2}}\sqrt{\gamma }\,d^{D-2}\varphi
\,r^{2n-1}f\int\limits_{0}^{1}dt\int\limits_{0}^{t}ds\,\delta _{\lbrack
n_{1}\cdots n_{2n-1}]}^{[m_{1}\cdots m_{2n-1}]}\mathcal{F}%
_{m_{4}m_{5}}^{n_{4}n_{5}}\cdots \mathcal{F}%
_{m_{2n-2}m_{2n-1}}^{n_{2n-2}n_{2n-1}}  \nonumber \\
&&\times \left[ \left( K_{\tau }^{\tau }\delta
_{m_{1}}^{n_{1}}+K_{m_{1}}^{n_{1}}\right) \mathcal{F}%
_{m_{2}m_{3}}^{n_{2}n_{3}}+2\left( n-1\right) \,K_{m_{1}}^{n_{1}}\delta
_{m_{2}}^{n_{2}}\left( -t^{2}K_{\tau }^{\tau }K_{m_{3}}^{n_{3}}+\frac{s^{2}}{%
\ell _{e\!f\!f}^{2}}\,\delta _{m_{3}}^{n_{3}}\right) \right] \,,
\end{eqnarray}%
or, more explicitly, after using the expressions (\ref{KBH}) and (\ref{calR}%
),%
\begin{eqnarray}
\int\limits_{\partial \mathcal{M}}d^{2n}x\,B_{2n}^{E} &=&2n\left(
2n-1\right) !\,\beta \,\text{Vol}(\Gamma _{2n-1})\,\lim_{r\rightarrow \infty
}\int\limits_{0}^{1}dt\int\limits_{0}^{t}ds\left( k-t^{2}f^{2}+s^{2}\,\frac{%
r^{2}}{\ell _{e\!f\!f}^{2}}\right) ^{n-2}\times  \nonumber \\
&&\left[ \frac{r(f^{2})^{\prime }}{2}\left( k-\left( 2n-1\right)
t^{2}f^{2}+s^{2}\frac{r^{2}}{\ell _{e\!f\!f}^{2}}\right) +f^{2}\left(
k-t^{2}f^{2}+\left( 2n-1\right) s^{2}\frac{r^{2}}{\ell _{e\!f\!f}^{2}}%
\right) \right] .
\end{eqnarray}%
In order to recognize the asymptotic charges from the Euclidean action, it
is especially convenient to convert the above double integrals into a
single-parameter integration, i.e.,
\begin{eqnarray*}
&&\int\limits_{0}^{1}dt\int\limits_{0}^{t}ds\,\left( k-\left( 2n-1\right)
t^{2}f^{2}+s^{2}\frac{r^{2}}{\ell _{e\!f\!f}^{2}}\right) \left(
k-t^{2}f^{2}+s^{2}\frac{r^{2}}{\ell _{e\!f\!f}^{2}}\right) ^{n-2} \\
&=&\int\limits_{0}^{1}dt\left( k-f^{2}+t^{2}\frac{r^{2}}{\ell _{e\!f\!f}^{2}}%
\right) ^{n-1}-\int\limits_{0}^{1}dt\,t\,\left( k-t^{2}f^{2}+t^{2}\frac{r^{2}%
}{\ell _{e\!f\!f}^{2}}\right) ^{n-1}
\end{eqnarray*}%
and%
\[
\int\limits_{0}^{1}dt\int\limits_{0}^{t}ds\,\left( k-t^{2}f^{2}+\left(
2n-1\right) s^{2}\frac{r^{2}}{\ell _{e\!f\!f}^{2}}\right) \left(
k-t^{2}f^{2}+s^{2}\frac{r^{2}}{\ell _{e\!f\!f}^{2}}\right)
^{n-2}=\int\limits_{0}^{1}dt\,t\,\left( k-t^{2}f^{2}+t^{2}\frac{r^{2}}{\ell
_{e\!f\!f}^{2}}\right) ^{n-1}.
\]%
Thus, the surface term is expressed as
\begin{eqnarray}
\int\limits_{\partial \mathcal{M}}d^{2n}x\,B_{2n}^{E} &=&n\left( 2n-1\right)
!\,\beta \text{Vol}(\Gamma _{2n-1})\lim_{r\rightarrow \infty }\left[
r^{2n-1}(f^{2})^{\prime }\int\limits_{0}^{1}dt\left( \frac{k-f^{2}}{r^{2}}+%
\frac{t^{2}}{\ell _{e\!f\!f}^{2}}\right) ^{n-1}\right. +  \nonumber \\
&&\qquad \qquad +\left. 2\left( f^{2}-\frac{r(f^{2})^{\prime }}{2}\right)
\int\limits_{0}^{1}dt\,t\,\left( k-t^{2}f^{2}+t^{2}\,\frac{r^{2}}{\ell
_{e\!f\!f}^{2}}\right) ^{n-1}\right] \,.
\end{eqnarray}

When the above boundary term is added to the bulk Euclidean action (\ref%
{Euclid_bulk}) with a suitable coupling constant,
\begin{equation}
I_{2n+1}^{E}=I_{0}^{E}+c_{2n}\int\limits_{\partial \mathcal{M}%
}d^{2n}x\,B_{2n}^{E}\,,
\end{equation}%
one gets%
\begin{eqnarray}
I_{2n+1}^{E} &=&\frac{\beta \,\text{Vol}(\Gamma _{2n-1})}{16\pi G}\,\left. %
\left[ r^{2n-1}(f^{2})^{\prime }\left( \rule{0in}{0.2in}1-2\alpha \left(
D-2\right) \left( D-3\right) \,\frac{f^{2}-k}{r^{2}}\right) \right]
\right\vert _{r_{+}}^{\infty }  \nonumber \\
&&+\beta \text{Vol}(\Gamma _{2n-1})\,nc_{2n}\,\left( 2n-1\right) !\left[
r^{2n-1}(f^{2})^{\prime }\int\limits_{0}^{1}dt\left( \frac{k-f^{2}}{r^{2}}+%
\frac{t^{2}}{\ell _{e\!f\!f}^{2}}\right) ^{n-1}\right.   \nonumber \\
&&\left. \left. +2\,\int\limits_{0}^{1}dt\,t\,\left( f^{2}-\frac{%
r(f^{2})^{\prime }}{2}\right) \left( k+t^{2}\left( \frac{r^{2}}{\ell
_{e\!f\!f}^{2}}-f^{2}\right) \right) ^{n-1}\right] \right\vert ^{\infty
}+\beta Q\Phi \,.  \label{Iodd3lines}
\end{eqnarray}%
Thus, the contribution coming from radial infinity in first two lines can be
identified with $\beta M$, where the mass is \cite{Miskovic-OleaNEDcharges}%
\begin{eqnarray}
M &=&\frac{\text{Vol}(\Gamma _{2n-1})}{16\pi G}\,\lim_{r\rightarrow \infty
}\,r^{2n-1}(f^{2})^{\prime }\left[ \rule{0pt}{22pt}1+2\alpha \left(
2n-1\right) \left( 2n-2\right) \,\frac{k-f^{2}}{r^{2}}+\right.   \nonumber \\
&&\hspace{-0.5cm}+\left. 16\pi
G(2n-1)!\,nc_{2n}\int\limits_{0}^{1}dt\,\left( \frac{k-f^{2}}{r^{2}}+\frac{%
t^{2}}{\ell _{e\!f\!f}^{2}}\right) ^{n-1}\right] ,
\end{eqnarray}%
which agrees with the mass formula (\ref{BH Mass}) when the metric function
is expanded. The term with parametric integration in the third line of Eq.(%
\ref{Iodd3lines}) is $\beta $ times the vacuum energy $E_{vac}$, which is
written in the form%
\begin{equation}
E_{vac}=2n\left( 2n-1\right) !c_{2n}\text{Vol}(\Gamma
_{2n-1})\,\lim_{r\rightarrow \infty }\,\int\limits_{0}^{1}dt\,t\left( f^{2}-%
\frac{r\left( f^{2}\right) ^{\prime }}{2}\right) \left[ k+\left( \frac{r^{2}%
}{\ell _{e\!f\!f}^{2}}-f^{2}\right) \,t^{2}\right] ^{n-1}.
\end{equation}

The consistency of the black hole thermodynamics is therefore verified
through the QSR, which involves the black hole entropy (\ref{S EGB})
\begin{equation}
G(T,\Phi )=T\,I_{2n+1}^{E}=U-T\,S+Q\Phi \,,
\end{equation}%
for a total energy which includes the zero-point energy $E_{vac}$,%
\begin{equation}
U=M+E_{vac}\,.
\end{equation}

\subsection{Thermodynamic charges}

Until now, we have obtained the QSR (\ref{QSR1}) for electrically charged
black holes in gravitating NED, which involves the total energy $U$ and the
electric charge $Q$ as asymptotic Noether charges, computed in Ref.\cite%
{Miskovic-OleaNEDcharges}, and the entropy $S$ as the Wald charge at the
horizon. However, strictly speaking, one should be able to reproduce these
quantities from thermodynamic relations in an independent way.

From that point of view, the gravitational entropy should be calculated from
the Gibbs free energy as
\begin{equation}
S_{TD}=-\left( \frac{\partial G}{\partial T}\right) _{\Phi }\,,
\label{entropy}
\end{equation}%
and the electric charge
\begin{equation}
Q_{TD}=\left( \frac{\partial G}{\partial \Phi }\right) _{T}\,,
\label{TD_charge}
\end{equation}%
whereas the internal energy can be derived as a thermodynamic quantity from
\begin{equation}
U_{TD}=G-T\left( \frac{\partial G}{\partial T}\right) _{\Phi }-\Phi \left(
\frac{\partial G}{\partial \Phi }\right) _{T}\,.  \label{internal}
\end{equation}

In general, arbitrary variations of $G$ in Eq.(\ref{Gdef}) produce in terms
of the Noether charges%
\begin{equation}
dG=dU-TdS+\Phi dQ-SdT+Qd\Phi \,.
\end{equation}%
It is clear that if Noether charges satisfy the First Law (\ref{First Law}),
then the relations (\ref{entropy}--\ref{internal}) identify thermodynamic
with Noether charges.

In order to prove the First Law, it is convenient to introduce the variable $%
\eta $,
\begin{equation}
\eta \equiv \frac{q}{r_{+}^{D-2}}\,,  \label{eta}
\end{equation}%
and write the electric charge as%
\begin{equation}
Q(r_{+},\eta )=4\text{Vol}(\Gamma _{D-2})\eta \,r_{+}^{D-2}\,.
\end{equation}

A more explicit expression for the Hawking temperature is%
\begin{equation}
T(r_{+},\eta )=\frac{\left( D-3\right) k+\frac{\left( D-1\right) \,r_{+}^{2}%
}{\ell ^{2}}+\alpha \,\left( D-3\right) \left( D-4\right) \left( D-5\right)
\,\frac{k^{2}}{r_{+}^{2}}+\frac{16\pi G\,r_{+}^{2}}{D-2}\,\left( \mathcal{L}%
_{+}-4\eta E_{+}\right) }{4\pi r_{+}\,\left( 1+2\alpha \,\left( D-3\right)
\left( D-4\right) \,\frac{k}{r_{+}^{2}}\right) \,}\,,  \label{T(r+,eta)}
\end{equation}%
where the index `$+$' denotes quantities evaluated at $r=r_{+}$.

The Generalized Gauss law,%
\begin{equation}
E_{+}\,\left. \frac{\partial \mathcal{L}}{\partial F^{2}}\right\vert
_{r_{+}}=-\eta \,,
\end{equation}%
determines the electric field $E_{+}=E(r_{+},q)$ such that it is a function
of $\eta $ only, and so are $\mathcal{L}_{+}$ and its first derivative. A
useful relation is
\begin{equation}
\frac{\partial \mathcal{L}_{+}}{\partial \eta }=\left. \frac{\partial
\mathcal{L}}{\partial F^{2}}\right\vert _{r_{+}}\frac{\partial F^{2}}{%
\partial \eta }=4\eta \,\frac{\partial E_{+}}{\partial \eta }\,.
\end{equation}

From the definition of horizon, $f(r_{+})=0$, we also obtain
\begin{equation}
U(r_{+},\eta )=E_{vac}+(D-2)\,\frac{\text{Vol}(\Gamma _{D-2})}{16\pi G}\left[
kr_{+}^{D-3}+k^{2}\alpha \left( D-3\right) \left( D-4\right) \,r_{+}^{D-5}+%
\frac{r_{+}^{D-1}}{\ell ^{2}}+\frac{16\pi G\,\mathcal{T}_{+}}{D-2}\right] \,,
\end{equation}%
where $E_{vac}$ is vanishing for even dimensions and $\mathcal{T}_{+}$ is
the function (\ref{I(q,r)}) evaluated at the horizon,
\begin{eqnarray}
\mathcal{T}_{+} &=&\frac{1}{D-1}\,\left. \left( \rule{0pt}{12pt}r^{D-1}%
\mathcal{L}-4qrE+\left( D-2\right) 4q\phi \right) \right\vert _{\infty
}^{r_{+}}  \nonumber \\
&=&\frac{1}{D-1}\,\left( r_{+}^{D-1}\mathcal{L}_{+}-4r_{+}^{D-1}\eta
E_{+}-\left( D-2\right) 4r_{+}^{D-2}\eta \Phi \right) \,.  \label{Tau+}
\end{eqnarray}%
Finally, the electric potential between infinity and the event horizon has
the form%
\begin{equation}
\Phi (r_{+},\eta )=-\frac{r_{+}\,\eta ^{\frac{1}{D-2}}}{D-2}%
\int\limits_{0}^{\eta }du\,u^{-\frac{D-1}{D-2}}\,E(r_{+},u)\,,
\end{equation}%
where $u=q/r^{D-2}=\eta \,\left( r_{+}/r\right) ^{D-2}$.

When varied, the entropy, charge and total energy change as
\begin{eqnarray}
dS &=&\frac{\text{Vol}(\Gamma _{D-2})}{4G}\,\left( D-2\right)
r_{+}^{D-3}\left( 1+2\alpha \left( D-3\right) \left( D-4\right) \,\frac{k}{%
r_{+}^{2}}\right) \,dr_{+}\,,  \label{dS} \\
dQ &=&4\text{Vol}(\Gamma _{D-2})r_{+}^{D-3}\left( \rule{0pt}{15pt}\left(
D-2\right) \eta \,dr_{+}+r_{+}\,d\eta \right) \,,  \label{dQ} \\
dU &=&(D-2)\,\frac{\text{Vol}(\Gamma _{D-2})}{16\pi G}\,r_{+}^{D-4}\left[
\left( D-3\right) k+\alpha \left( D-3\right) \left( D-4\right) \left(
D-5\right) \,\frac{k^{2}}{r_{+}^{2}}\right.  \nonumber \\
&&\qquad +\left. \left( D-1\right) \,\frac{r_{+}^{2}}{\ell ^{2}}\right]
dr_{+}+\text{Vol}(\Gamma _{D-2})\,d\mathcal{T}_{+}\,.
\end{eqnarray}

The last line involves the variation of $\mathcal{T}_{+}$ which, using Eq.(%
\ref{Tau+}) and the variation of $\Phi $,%
\begin{equation}
d\Phi =\frac{\Phi }{r_{+}}\,dr_{+}+\frac{\Phi -r_{+}E_{+}}{\left( D-2\right)
\eta }\,d\eta \,,
\end{equation}%
can be left in the more suitable form
\begin{equation}
d\mathcal{T}_{+}=\left[ \rule{0pt}{15pt}r_{+}^{D-2}\left( \mathcal{L}%
_{+}-4\eta E_{+}\right) -4\left( D-2\right) r_{+}^{D-3}\eta \Phi \right]
dr_{+}-4r_{+}^{D-2}\Phi \,d\eta \,.
\end{equation}
With the formula for $T$ given by Eq.(\ref{T(r+,eta)}), we finally find that
\begin{eqnarray}
dU &=&T\,(D-2)\,\frac{\text{Vol}(\Gamma _{D-2})}{4G}\,r_{+}^{D-3}\left(
1+2\alpha \,\left( D-3\right) \left( D-4\right) \,\frac{k}{r_{+}^{2}}\right)
dr_{+}  \nonumber \\
&&-4\text{Vol}(\Gamma _{D-2})r_{+}^{D-3}\,\left( \rule{0pt}{15pt}\left(
D-2\right) \eta \,dr_{+}+r_{+}\,d\eta \right) \Phi \,,
\end{eqnarray}%
from where we recognize the variations $dS$ and $dQ$ as in Eqs. (\ref{dS})
and (\ref{dQ}), such that the First Law holds.

In this case, the Gibbs free energy satisfies Eq.(\ref{dG}) and, as a
consequence, the thermodynamic relations (\ref{entropy}--\ref{internal}) are
also valid for the corresponding Noether charges, what implies%
\begin{eqnarray}
S_{TD} &=&S\,,  \nonumber \\
Q_{TD} &=&Q\,,  \nonumber \\
U_{TD} &=&U=M+E_{vac}\,.
\end{eqnarray}%
It is worthwhile stressing that, while the proof of First Law is insensitive
to the presence of the vacuum energy $E_{vac}$, the last equation defines
the internal energy as the total energy of the black hole, as expected in
the context of AdS/CFT correspondence.

\section{Conclusions}

In this paper, we have studied the thermodynamics of topological static AdS
black holes in Einstein-Gauss-Bonnet theory in the presence of an electric
field described by an arbitrary nonlinear electrodynamics. This is done
using a background-independent regularization scheme which considers the
addition of counterterms depending on both the intrinsic and extrinsic
curvature of the boundary metric, to the action.

The fact that the Kounterterm series is given as a compact expression, i.e.,
in a parametric representation of the polynomials, provides us with a
practical tool to evaluate the regularized action in all dimensions.
Likewise, this also makes easier recognizing the quantities appearing at
radial infinity as the conserved quantities of the theory. This issue is
particularly relevant, because it connects the finiteness of the Euclidean
action with the problem of regularization of the asymptotic charges treated
in Ref.\cite{Miskovic-OleaNEDcharges} and, ultimately, with a well-posed
variational principle.

In the standard counterterms method, it is not possible to separate the
contributions to the quasilocal stress tensor that produce the black hole
mass $M$ from those which give rise to the vacuum energy. This means that in
the evaluation of the Euclidean action $I^{E}$ one cannot clearly identified
the term $\beta E_{vac}$ from the terms at $r=\infty $ in all odd
dimensions. Nonetheless, one could work out a relation between the
Kounterterms and the intrinsic prescription given by holographic
renormalization, by taking the asymptotic expansion of the extrinsic
curvature (for a similar procedure in Einstein-Hilbert gravity, see Ref.\cite%
{Miskovic-Olea 4D}).

We also performed an independent checking, which is obtaining the total
energy $U$ and electric charge $Q$ as thermodynamic quantities from Eqs.(\ref%
{internal}) and (\ref{TD_charge}), respectively. As expected, the internal
energy contains a Casimir contribution in odd dimensions, consisting with
the Noether approach.

The generality of the procedure is a consequence of the fact that, in the
context of Kounterterm regularization, the Quantum Statistical Relation
appears as a thermodynamic identity of the gravity-NED action\ plus boundary
terms. This suggests an extension of the above results to other Lovelock
theories and, probably, to other matter couplings, as long as one can define
the asymptotic behavior in terms of the curvature as in Eq.(\ref{LAdSleff})
(for a thermodynamic study of higher derivative corrections in the Abelian
gauge field of the form $F^{4}$, see Ref.\cite{Anninos-Pastras}).

\section*{Acknowledgments}

This work was funded by FONDECYT Grants 11070146, 1090357 and 1100755. O.M.
is supported in part by Project MECESUP UCV0602 and the PUCV through the
projects 123.797/2007 and 123.705/2010.

\appendix

\section{Useful identities \ \label{Delta}}

The totally-antisymmetric Kronecker delta of rank $p$ is defined as the
determinant
\begin{equation}
\delta _{\left[ \mu _{1}\cdots \mu _{p}\right] }^{\left[ \nu _{1}\cdots \nu
_{p}\right] }:=\left\vert
\begin{array}{cccc}
\delta _{\mu _{1}}^{\nu _{1}} & \delta _{\mu _{1}}^{\nu _{2}} & \cdots &
\delta _{\mu _{1}}^{\nu _{p}} \\
\delta _{\mu _{2}}^{\nu _{1}} & \delta _{\mu _{2}}^{\nu _{2}} &  & \delta
_{\mu _{2}}^{\nu _{p}} \\
\vdots &  & \ddots &  \\
\delta _{\mu _{p}}^{\nu _{1}} & \delta _{\mu _{p}}^{\nu _{2}} & \cdots &
\delta _{\mu _{p}}^{\nu _{p}}%
\end{array}%
\right\vert \,.
\end{equation}%
A contraction of $k\leq p$ indices in the Kronecker delta of rank $p$
produces a delta of rank $p-k$,
\begin{equation}
\delta _{\left[ \mu _{1}\cdots \mu _{k}\cdots \mu _{p}\right] }^{\left[ \nu
_{1}\cdots \nu _{k}\cdots \nu _{p}\right] }\,\delta _{\nu _{1}}^{\mu
_{1}}\cdots \delta _{\nu _{k}}^{\mu _{k}}=\frac{\left( N-p+k\right) !}{%
\left( N-p\right) !}\,\delta _{\left[ \mu _{k+1}\cdots \mu _{p}\right] }^{%
\left[ \nu _{k+1}\cdots \nu _{p}\right] }\,,
\end{equation}%
where $N$ is the range of indices.

Using this compact notation, the Einstein tensor (\ref{Einstein tensor}) can
be rewritten in terms of the AdS radius as
\begin{equation}
G_{\nu }^{\mu }=-\frac{1}{2}\,\delta _{\lbrack \nu \nu _{1}\nu _{2}]}^{[\mu
\mu _{1}\mu _{2}]}\left( \frac{1}{2}R_{\mu _{1}\mu _{2}}^{\nu _{1}\nu _{2}}+%
\frac{1}{\ell ^{2}}\,\delta _{\mu _{1}}^{\nu _{1}}\delta _{\mu _{2}}^{\nu
_{2}}\right) \,,  \label{Einstein deltas}
\end{equation}%
and, in a similar fashion, the Laczos tensor (\ref{Lanczos tensor}) adopts
the form%
\begin{equation}
H_{\nu }^{\mu }=-\frac{\alpha }{8}\,\delta _{\lbrack \nu \nu _{1}\cdots \nu
_{4}]}^{[\mu \mu _{1}\cdots \mu _{4}]}\,R_{\mu _{1}\mu _{2}}^{\nu _{1}\nu
_{2}}R_{\mu _{3}\mu _{4}}^{\nu _{3}\nu _{4}}\,.  \label{Lanczos deltas}
\end{equation}%
In order to identify the equation of motion for the metric in the evaluation
of the Euclidean action, it is convenient to employ the above relations to
convert (\ref{E^mu_nu}) into%
\begin{eqnarray}
\mathcal{E}_{\nu }^{\mu } &=&-\frac{1}{2\left( D-3\right) \left( D-4\right) }%
\,\delta _{\lbrack \nu \nu _{1}\cdots \nu _{4]}}^{[\mu \mu _{1}\cdots \mu
_{4}]}\,\left[ \frac{1}{2}\,R_{\mu _{1}\mu _{2}}^{\nu _{1}\nu _{2}}\,\delta
_{\mu _{3}}^{\nu _{3}}\delta _{\mu _{4}}^{\nu _{4}}+\right.  \nonumber \\
&&\qquad \left. +\frac{1}{\ell ^{2}}\,\delta _{\mu _{1}}^{\nu _{1}}\delta
_{\mu _{2}}^{\nu _{2}}\,\delta _{\mu _{3}}^{\nu _{3}}\delta _{\mu _{4}}^{\nu
_{4}}+\frac{\alpha }{4}\,\left( D-3\right) \left( D-4\right) \,R_{\mu
_{1}\mu _{2}}^{\nu _{1}\nu _{2}}R_{\mu _{3}\mu _{4}}^{\nu _{3}\nu _{4}}%
\right] -8\pi G\,T_{\nu }^{\mu }\,.  \label{Emu_nu}
\end{eqnarray}

\section{Gauss-normal coordinate frame \label{Gauss-normalApp}}

In the Gauss-normal coordinate system (\ref{Gaussian}), the only relevant
components of the connection $\Gamma _{\mu \nu }^{\alpha }$ are expressed in
terms of the extrinsic curvature $K_{ij}=-\frac{1}{2N}\,h_{ij}^{\prime }$ as
\begin{equation}
\Gamma _{ij}^{r}=\frac{1}{N}\,K_{ij\,},\qquad \Gamma
_{rj}^{i}=-NK_{j}^{i}\,,\qquad \Gamma _{rr}^{r}=\frac{N^{\prime }}{N}\,.
\label{KChr}
\end{equation}%
The radial foliation (\ref{Gaussian}) implies the Gauss-Codazzi relations
for the spacetime curvature, as well,
\begin{eqnarray}
R_{kl}^{ir} &=&\frac{1}{N}\,\left( \nabla _{l}K_{k}^{i}-\nabla
_{k}K_{l}^{i}\right) \,,  \label{Codazzi2} \\
R_{kr}^{ir} &=&\frac{1}{N}\,\left( K_{k}^{i}\right) ^{\prime
}-K_{l}^{i}\,K_{k}^{l}\,,  \label{Codazzi3} \\
R_{kl}^{ij} &=&\mathcal{R}_{kl}^{ij}(h)-K_{k}^{i}\,K_{l}^{j}+K_{l}^{i}%
\,K_{k}^{j}\,\,,  \label{Codazzi1}
\end{eqnarray}%
where $\nabla _{i}=\nabla _{i}(h)$ is the covariant derivative defined in
the Christoffel symbol of the boundary $\Gamma _{ij}^{k}(g)=\Gamma
_{ij}^{k}(h)$ and $\mathcal{R}_{kl}^{ij}(h)$ is the intrinsic curvature of
the boundary.

\section{Topological black hole metric \label{TBH metric}}

In the static topological black hole ansatz (\ref{BH}), the extrinsic
curvature takes the form%
\begin{equation}
K_{j}^{i}=-\frac{1}{2N}\,h^{ik}h_{kj}^{\prime }=\left(
\begin{array}{cc}
-f^{\prime } & 0 \\
0 & -\frac{f}{r}\,\delta _{n}^{m}%
\end{array}%
\right) \,,  \label{KBH}
\end{equation}%
where prime denotes radial derivative, and the non-vanishing components of
the boundary curvature are%
\begin{equation}
\mathcal{R}_{m_{1}m_{2}}^{n_{1}n_{2}}(h)=\frac{k}{r^{2}}\,\delta _{\lbrack
m_{1}m_{2}]}^{[n_{1}n_{2}]}\,.  \label{calR}
\end{equation}

The spacetime Riemann tensor $R_{\lambda \rho }^{\mu \nu }$ is then given by%
\begin{eqnarray}
R_{tr}^{tr} &=&-\frac{1}{2}\,\left( f^{2}\right) ^{\prime \prime }\,,
\nonumber \\
R_{tm}^{tn} &=&R_{rm}^{rn}=-\frac{1}{2r}\,\left( f^{2}\right) ^{\prime
}\,\delta _{m}^{n}\,,  \label{RiemannBH} \\
R_{kl}^{mn} &=&\frac{1}{r^{2}}\,\left( k-f^{2}\right) \,\delta _{\lbrack
kl]}^{[mn]}\,.  \nonumber
\end{eqnarray}


\begin{thebibliography}{99}

\bibitem{AdS/CFT} J.M. Maldacena, \emph{The large }$N$\emph{\ limit of
superconformal field theories}, Adv. Theor. Math. Phys. \textbf{2}, 231
(1998); Int. J. Theor. Phys. \textbf{38}, 1113 (1999). [hep-th/9711200 ];
S.S. Gubser, I.R. Klebanov and A.M. Polyakov, \emph{A semiclassical limit of
the gauge string correspondence}, Nucl. Phys. \textbf{B636}, 99 (2002).
[hep-th/0204051]; E. Witten,\emph{\ Anti-de Sitter space and holography},
Adv. Theor. Math. Phys. \textbf{2}, 253 (1998). [hep-th/9802150]

\bibitem{KSS} P. Kovtun, D. T. Son and A. O. Starinets, \emph{Holography and
hydrodynamics: diffusion on stretched horizons}, J. High Energy Phys.
\textbf{10}, 064 (2003). [hep-th/0309213]

\bibitem{viscosity} M. Brigante, H. Liu, R. C. Myers, S. Shenker and S.
Yaida, \emph{Viscosity bound violation in higher derivative gravity}, Phys.
Rev. \textbf{D77}: 126006 (2008). [arXiv:0712.0805]; Y. Kats and P. Petrov,
\emph{Effect of curvature squared corrections in AdS on the viscosity of the
dual gauge theory}, J. High Energy Phys. \textbf{01}: 044 (2009).
[arXiv:0712.0743].
\bibitem{Cai-Sun} R-G. Cai and Y-W. Sun, \emph{Shear viscosity from AdS
Born-Infeld black holes}, J. High Energy Phys. \textbf{09}: 115 (2008).
[arXiv:0807.2377]

\bibitem{Ge-Matsuo-Shu-Sin-Tsukioka} X.H. Ge, Y. Matsuo, F.W. Shu, S.J. Sin
and T. Tsukioka, \emph{Viscosity bound, causality violation and instability
with stringy correction and charge}, J. High Energy Phys. \textbf{10}, 009
(2008). [arXiv:0808.2354]



\bibitem{Gregory-Kanno-Soda} R. Gregory, S. Kanno and J. Soda, \emph{%
Holographic superconductors with higher curvature corrections}, J. High
Energy Phys. \textbf{10}: 010 (2009). [arXiv:0907.3203]

\bibitem{Pan-Wang-Papantonopoulos-Oliveira-Pavan} Q. Y. Pan, B. Wang, E.
Papantonopoulos, J. Oliveira and A. Pavan, \emph{Holographic superconductors
with various condensates in Einstein-Gauss-Bonnet gravity}, Phys. Rev.
\textbf{D81}: 106007 (2010). [arXiv:0912.2475]

\bibitem{Jing-Chen} J. Jing and S. Chen, \emph{Holographic superconductors
in the Born-Infeld electrodynamics}, Phys. Lett. \textbf{B686}: 68 (2010).
[arXiv:1001.4227]; J. Jing, L. Wang, Q. Pan and S. Chen, \emph{Holographic
superconductors in Gauss-Bonnet gravity with Born-Infeld electrodynamics}.
[arXiv:1012.0644]

\bibitem{Gibbons-Perry-Pope} G.W. Gibbons, M.J. Perry and C.N. Pope, \emph{%
The First law of thermodynamics for Kerr-anti-de Sitter black holes}, Class.
Quant. Grav. \textbf{22}, 1503 (2005). [hep-th/0408217]


\bibitem{footnote1} Following the arguments given in Ref.\cite%
{Gibbons-Perry-Pope}, we adopt this nomenclature for this thermodynamic
formula.

\bibitem{footnote2} This fact is a consequence of breaking $SO(2n,2)$
invariance by the action of radial diffeomorphisms.

\bibitem{Balasubramanian-Kraus} V. Balasubramanian and P. Kraus, \emph{A
stress tensor for anti-de Sitter gravity}, Commun. Math. Phys. \textbf{208},
413 (1999). [hep-th/9902121]

\bibitem{Emparan-Johnson-Myers} R. Emparan, C.V. Johnson and R.C. Myers,
\emph{Surface terms as counterterms in the AdS/CFT correspondence},
Phys.Rev. D \textbf{60}, 104001 (1999). [hep-th/9903238]; R.B. Mann, \emph{%
Misner string entropy}, Phys. Rev. D \textbf{60}, 1040047 (1999).
[hep-th/9903229]

\bibitem{Henningson-Skenderis} M. Henningson and K. Skenderis, \emph{The
holographic Weyl anomaly}, J. High Energy Phys. \textbf{07}, 023 (1998).
[hep-th/9806087]

\bibitem{deHaro-Skenderis-Solodukhin} S. de Haro, K. Skenderis and S.
Solodukhin, \emph{Holographic reconstruction of space-time and
renormalization in the AdS/CFT correspondence}, Commun. Math. Phys. \textbf{%
217}, 595 (2001). [hep-th/0002230]

\bibitem{Brihaye-Radu} Y. Brihaye and E. Radu, \emph{Black objects in the
Einstein-Gauss-Bonnet theory with negative cosmological constant and the
boundary counterterm method}, J. High Energy Phys. \textbf{09}, 006 (2008).
[arXiv:0806.1396]

\bibitem{Liu-Sabra} J.T. Liu and W.A. Sabra, \emph{Hamilton-Jacobi
counterterms for Einstein-Gauss-Bonnet gravity}, Class. Quant. Grav. \textbf{%
27}, 175014 (2010). [arXiv:0807.1256]

\bibitem{Kofinas-Olea-GB} G. Kofinas and R. Olea, \emph{Vacuum energy in
Einstein-Gauss-Bonnet AdS gravity}, Phys. Rev. \textbf{D74}, 084035 (2006).
[hep-th/0606253]

\bibitem{Miskovic-OleaBI} O. Miskovic and R. Olea, \emph{%
Thermodynamics of Einstein-Born-Infeld black holes with negative
cosmological constant}, Phys. Rev. \textbf{D77}, 124048 (2008).
[arXiv:0802.2081]

\bibitem{Heisenberg-Euler} W. Heisenberg and H. Euler, \emph{Folgerungen aus
der Diracschen Theorie des Positrons}, Z. Phys. \textbf{98}, 714 (1936).
Translation by W. Korolevski and H. Kleinert, \emph{Consequences of Dirac's
Theory of the Positron}. [physics/0605038]

\bibitem{Born-Infeld} M. Born and I. Infeld, \emph{Foundations of the new
field theory}, Proc. R. Soc. \textbf{A144}, 425 (1934).

\bibitem{Fradkin-Tseytlin} E.S. Fradkin and A.A. Tseytlin, \emph{Nonlinear
electrodynamics from quantized strings}, Phys. Lett. \textbf{B163}, 123
(1985).

\bibitem{Hoffman-Gibbons-Rasheed} B. Hoffmann, \emph{Gravitational and
electromagnetic mass in the Born-Infeld electrodynamics}, Phys. Rev. \textbf{%
47}, 877 (1935); G. W. Gibbons and D. A. Rasheed, \emph{Electric-magnetic duality
rotations in nonlinear electrodynamics}, Nucl. Phys. \textbf{B454}, 185
(1995).

\bibitem{Oliveira} H.P. de Oliveira, \emph{Nonlinear charged black holes},
Class. Quant. Grav. \textbf{11}, 1469 (1994).

\bibitem{Soleng} H.H. Soleng, \emph{Charged black points in general
relativity coupled to the logarithmic }$U(1)$\emph{\ gauge theory}, Phys.
Rev. \textbf{D52}, 6178 (1995). [hep-th/9509033]

\bibitem{Maeda-Hassaine-Martinez} H. Maeda, M. Hassaine and C. Martinez,
\emph{Lovelock black holes with a nonlinear Maxwell field}, Phys. Rev.
\textbf{D79}, 044012 (2009). [arXiv:0812.2038]

\bibitem{Boulware-Deser} D. Boulware and S. Deser, \emph{String generated
gravity models}, Phys. Rev. Lett. \textbf{55}, 2656 (1985).

\bibitem{Miskovic-OleaNEDcharges} O. Miskovic and R. Olea, \emph{%
Conserved charges for black holes in Einstein-Gauss-Bonnet gravity coupled
to nonlinear electrodynamics in AdS space}, Phys. Rev. \textbf{D83}: 024011
(2011). [arXiv:1009.5763]


\bibitem{Cai} R-G. Cai, \emph{Gauss-Bonnet black holes
in AdS spaces}, Phys. Rev. \textbf{D65}: 084014 (2002). [hep-th/0109133]

\bibitem{Cvetic-Nojiri-Odintsov} M. Cvetic, S. Nojiri and S.D. Odintsov,
\emph{Black hole thermodynamics and negative entropy in de Sitter
and anti-de Sitter Einstein-Gauss-Bonnet gravity}
Nucl. Phys. \textbf{B628}, 295 (2002). [hep-th/0112045]

\bibitem{Deser-Tekin} S. Deser and B. Tekin, \emph{Energy in generic higher
curvature gravity theories}, Phys. Rev. \textbf{D67}: 084009 (2003).
[hep-th/0212292]

\bibitem{Padilla} A. Padilla, \emph{Surface terms and the Gauss-Bonnet
Hamiltonian}, Class. Quant. Grav. \textbf{20}: 3129 (2003). [gr-qc/0303082]

\bibitem{Gravanis} E. Gravanis, \emph{Conserved charges in (Lovelock)
gravity in first order formalism}, Phys. Rev. \textbf{D81}: 084013 (2010).
[arXiv:1004.3582]

\bibitem{Kastor}  For a discussion on Komar formula for Lovelock gravity
theories see, e.g., D. Kastor, \emph{Komar integrals in higher (and lower)
derivative gravity}, Class. Quant. Grav. \textbf{25}: 175007
(2008). [arXiv:0804.1832]


\bibitem{Iyer-Wald} V. Iyer and R.M. Wald, \emph{Some properties of Noether
charge and a proposal for dynamical black hole entropy}, Phys. Rev. \textbf{%
D50}, 846 (1994). [gr-qc/9403028]

\bibitem{Clunan-Ross-Smith} T. Clunan, S.F. Ross and D.J. Smith, \emph{On
Gauss-Bonnet black hole entropy}, Class. Quant. Grav. \textbf{21},
3447 (2004). [gr-qc/0402044]

\bibitem{Paranjape-Sarkar-Padmanabhan} A. Paranjape, S. Sarkar and T.
Padmanabhan, \emph{Thermodynamic route to field equations in Lancos-Lovelock
gravity}, Phys. Rev. \textbf{D74}: 104015 (2006). [hep-th/0607240]

\bibitem{footnote3} The reader might be confused with the unusual dependence
on the Christoffel symbol of Eq.(\ref{I-W tensor}) as compared with the
formulas appearing in the literature. However, it is straightforward to
prove that the charge can be equivalently written in terms of the covariant
derivative of the vector $\xi ^{\mu }$.

\bibitem{Kofinas-Olea} G. Kofinas and R. Olea, \emph{Universal
regularization prescription for Lovelock AdS gravity}, J. High Energy Phys.
\textbf{11}, 069 (2007). [arXiv: 0708.0782]; \emph{Universal Kounterterms in
Lovelock AdS gravity}, Fortsch. Phys. \textbf{56}, 957 (2008).
[arXiv:0806.1197]

\bibitem{Olea-K} R. Olea, \emph{Regularization of odd-dimensional AdS
gravity: Kounterterms}, J. High Energy Phys. \textbf{04}, 073 (2007).
[hep-th/0610230]

\bibitem{Olea-KerrBH} R. Olea, \emph{Mass, angular momentum and
thermodynamics in four-dimensional Kerr-AdS black holes}, J. High Energy
Phys. \textbf{06}, 023 (2005). [hep-th/0504233]

\bibitem{Miskovic-Olea 4D} O. Miskovic and R. Olea, \emph{%
Topological regularization and self-duality in four-dimensional anti-de
Sitter gravity}, Phys. Rev. \textbf{D79}, 124020 (2009). [arXiv:0902.2082]

\bibitem{Anninos-Pastras} D. Anninos and G. Pastras,
 \emph{Thermodynamics of the Maxwell-Gauss-Bonnet anti-de Sitter black
 hole with higher derivative gauge corrections}, J. High Energy Phys.
\textbf{07}, 030 (2009). [arXiv:0807.3478]

\end{thebibliography}
\end{document}